\documentclass[runningheads]{llncs}
\usepackage{graphicx}
\usepackage{hyperref}
\usepackage{amssymb}
\usepackage{booktabs}
\usepackage{color, colortbl}
\usepackage[dvipsnames]{xcolor}
\usepackage{stmaryrd}
\usepackage{amsmath}
\usepackage{multirow}

\definecolor{aliceblue}{rgb}{0.94, 0.97, 1.0}

\begin{document}
\newcommand{\LNCI}{\texttt{LNC1}}
\newcommand{\LNCII}{\texttt{LNC2}}
\newcommand{\TFCI}{\texttt{TFC1}}
\newcommand{\TFCII}{\texttt{TFC2}}
\newcommand{\MTDC}{\texttt{M-TDC}}
\newcommand{\STMCI}{\texttt{STMC1}}
\newcommand{\STMCII}{\texttt{STMC2}}
\newcommand{\STMCIII}{\texttt{STMC3}}
\newcommand{\LBI}{\texttt{LB1}}
\newcommand{\LBII}{\texttt{LB2}}
\newcommand{\TP}{\texttt{TP}}
\newcommand{\bert}{BERT}
\newcommand{\robust}{Robust04}
\graphicspath{ {./plots/} }

\newcommand\todo[1]{\textcolor{red}{$\triangleright$ #1 $\triangleleft$}}

\title{Diagnosing BERT with Retrieval Heuristics}

\author{Arthur C\^{a}mara and Claudia Hauff}
\institute{Delft University of Technology\\Delft, the Netherlands\\\{a.barbosacamara,c.hauff\}@tudelft.nl}

\maketitle             

\begin{abstract}
Word embeddings, made widely popular in 2013 with the release of word2vec, have become a mainstay of NLP engineering pipelines. Recently, with the release of \bert{}, word embeddings have moved from the term-based embedding space to the contextual embedding space---each term is no longer represented by a single low-dimensional vector but instead each term and \emph{its context} determine the vector weights. \bert{}'s setup and architecture have been shown to be general enough to be applicable to many natural language tasks. Importantly for Information Retrieval (IR), in contrast to prior deep learning solutions to IR problems which required significant tuning of neural net architectures and training regimes, ``vanilla \bert{}'' has been shown to outperform existing retrieval algorithms by a wide margin, including on tasks and corpora that have long resisted retrieval effectiveness gains over traditional IR baselines (such as \robust{}). In this paper, we employ the recently proposed axiomatic dataset analysis technique---that is, we create diagnostic datasets that each fulfil a retrieval heuristic (both term matching and semantic-based)---to explore what \bert{} is able to learn. In contrast to our expectations, we find \bert{}, when applied to a recently released large-scale web corpus with ad-hoc topics, to \emph{not} adhere to any of the explored axioms. At the same time, \bert{} outperforms the traditional query likelihood retrieval model by 40\%. This means that the axiomatic approach to IR (and its extension of diagnostic datasets created for retrieval heuristics) may in its current form not be applicable to large-scale corpora. Additional---different---axioms are needed.


\end{abstract}

\section{Introduction}

    Over the course of the past few years, IR has seen the introduction of a large number of successful deep learning approaches for solving all kinds of tasks previously tackled with hand-crafted features (within the learning to rank framework) or traditional retrieval models such BM25. 
    
    In 2017, with the introduction of the transformer architecture~\cite{Vaswani17NIPS}, a second wave of neural architectures for NLP has emerged. Approaches (and respective models) like \bert{}~\cite{DevlinCLT19NAACL}, XLNet~\cite{YangDYCSL19CoRR} and GPT-2~\cite{radford2019language} have shown that it is indeed possible for one general architecture to achieve state-of-the-art performance across very different NLP tasks (some of which are also related to IR tasks, such as question answering, reading comprehension, etc.).
        
    
    Ad-hoc retrieval, the task of ranking a set of documents given a single query, has long resisted the success of neural approaches, especially when employed across standard IR test collections such as \robust{}\footnote{\robust{} is a test collection employed at the TREC 2004 robust retrieval task~\cite{voorhees2004overview}, consisting of 528K newswire documents, 250 topics and 311K relevance judgements.}, which come with hundreds of topics (and thus relatively little training data). Often, the proposed neural approaches require a very careful design of their architecture. Also, the training regime and the input data transformations have to be \emph{just right}~\cite{Lin18SIGIRF} in order to beat or come close to well-tuned traditional IR baselines such as RM3~\cite{LavrenkoC01,abdul2004umass}\footnote{We want to emphasise here that this observation is specific to IR corpora with few training topics; for the very few corpora with hundreds of thousands of released topics (such as MSMarco) this observation does not hold.}. With the introduction of \bert{} in late 2018, this has finally changed.  Recently, a range of \bert{}-inspired approaches have been shown to clearly surpass all strong IR baselines on \robust{}~\cite{DaiC19SIGIR,YangZL2019Arxiv} and other IR corpora.
    
    It is still an open question though what exactly makes \bert{} and similar approaches perform so well on IR tasks. While recent works try to understand what \bert{}~learns most often by analysing attention values, e.g.,~\cite{TenneyDP2019ACL,NivenK19ACL,FernandoSA19SIGIR,ClarkKLD2019ArXiv}, analysing \bert{} under the IR light requires a different set of tools. While most NLP tasks optimise for precision, recall or other objective metrics, the goal of ad-hoc retrieval is to optimise for \emph{relevance}, a complex multidimensional and somewhat subjective concept~\cite{borlund2003concept}.
    
    In this paper, we set out to explore \bert{} under the IR lens, employing the concept of  \emph{diagnostic datasets}, an IR model analysis approach (inspired by similar NLP and computer vision approaches) proposed last year by Rennings et al.~\cite{RenningsMH19ECIR}. The idea behind these datasets is simple: each dataset is designed to fulfil one \emph{retrieval axiom}~\cite{FangTZ04SIGIR}, i.e., a heuristic that a good retrieval function should fulfil\footnote{As a concrete example, consider the TFC1~\cite{FangTZ04SIGIR} heuristic: \emph{The more occurrences of a query term a document has, the higher its retrieval score}.}. Each dataset contains query-documents instances (most often, a query and two documents) that the investigated model should rank in the correct order as determined by the heuristic. The extent to which a model correctly ranks those instances is allowing us to gain insights into what type of information the retrieval model pays attention to (or not) when ranking documents. While traditional retrieval models such as BM25~\cite{robertson2009probabilistic} can be analysed analytically, neural nets with their millions or even billions of learnt weights can only be analysed in such an empirical manner.

    More concretely, we attempt to analyse a version of \bert{}, Distil\bert{} (that was shown to attain 97\% of ``vanilla" \bert{} performance~\cite{sanh2019distilbert}), fine-tuned on the TREC 2019 Deep Learning track dataset\footnote{\url{https://microsoft.github.io/TREC-2019-Deep-Learning/}}. We extend previous work~\cite{RenningsMH19ECIR} by incorporating additional axioms (moving from term matching to semantic axioms). We find that  Distil\bert{} to outperform the traditional query likelihood (QL) model by 40\%. In contrast to our expectations however, we find that \bert{} does not adhere to any of the axioms we incorporate in our work. This implies that the currently existing axioms are \emph{not sufficient} and \emph{not applicable} to capture the heuristics that a strong supervised model learns (at least for the corpus and model we explore); it is not yet clear to what extent those results generalise beyond our model and corpus combination but it opens up a number of questions about the axiomatic approach to IR.

\section{Related Work}\label{sec:related}

\paragraph{Axiomatic Information Retrieval}
The use of axioms (or ``retrieval heuristics'') as a means to improve and understand information retrieval techniques is well established. It is an analytic technique to explore retrieval models and how best to improve them. In their seminal work, Fang et al.~\cite{FangTZ04SIGIR,FangZ05SIGIR} introduced a number of term-matching based \textit{heuristics} that models should follow in order to be successful in retrieval tasks. 
Subsequently, Fang et al.~\cite{Fang06SIGIR} proposed a set of axioms based on semantic matching and thus allowing non-exact matches to be accounted for in axiomatic retrieval. We apply these axioms in our work---albeit in a slightly adapted manner. Other applications for axioms in IR include document re-ranking based on a Learning to Rank scenario~\cite{HagenVGS16CIKM} and query expansion~\cite{Fang08ACL} by exploring similar axioms. It should be noted, that---while sensible---it cannot be assumed that these axioms are a good fit for all kinds of corpora; they represent a general notion of how a good retrieval function should behave. Recently, Rennings et al.~\cite{RenningsMH19ECIR} introduced \emph{diagnostic datasets} extracted from actual corpora that each fulfil one axiom. In contrast to the axiomatic approach, which requires an analytical evaluation of the retrieval functions under investigation, a diagnostic corpus enables us to analyse models' axiomatic performance that are too large to be analysed analytically (such as neural models with millions or even billions of parameters\footnote{As a concrete example, our \bert{} model contains 66 million parameters.}). Our work continues in that direction with a larger number of axioms (9 vs. 4) and the analysis of the current neural state-of-the-art (i.e., \bert{}).

\paragraph{Neural IR Models} Neural IR models, i.e., deep learning based approaches that tackle IR problems, have seen a massive rise in popularity in the last few years, with considerable success~\cite{MitraC18FTIR}. Models like DRMM~\cite{McDonaldBA18EMNLP}, ARCII~\cite{HuLLC14NIPS} and aNMM~\cite{YangAGC16CIKM} have been shown to be suitable for a range of IR tasks when sufficient training data is available; it remains at best unclear at smaller data scale whether the reported successes are not just an artefact of weak baselines~\cite{Lin18SIGIRF}.

Recently, a new wave of approaches, based on the transformer architecture~\cite{Vaswani17NIPS} has shown that, finally, neural models can significantly outperform traditional and well-tuned retrieval methods such as RM3~\cite{abdul2004umass}. Yang et al.~\cite{YangZL2019Arxiv} have shown that \bert{}, fine-tuned on the available TREC microblog datasets, and combined with a traditional retrieval approach such as query likelihood significantly outperforms well-tuned baselines, even on \robust{} which has shown to be a notoriously difficult dataset for neural models to do well on. With similar success, Dai and Callan~\cite{DaiC19SIGIR} have recently employed another \bert{} variant on~\robust{} and ClueWeb09. Lastly we point out, that works are now also beginning to appear, e.g.,~\cite{MacAvaneyYCG19SIGIR}, that use the contextual word embeddings produced by \bert{} in combination with another strong neural model, again with strong improvements over the existing baselines.

\paragraph{Analysing Neural IR Models}
As we aim to analyse \bert{}, we also consider how others have tackled this problem. Analysing neural models---whether for IR, NLP or another research domain---is not a trivial task. By now a great number of works have tried to light up the black box of the neural learning models~\cite{blackbox18EMNLP}, with varying degrees of success. Within IR, Pang et al.~\cite{PangLG0C17CoRR} have aimed to paint a complete and high-level picture of the neural IR area, comparing the behaviour of different approaches, and showing that interaction and representation-based models focus on different characteristics of queries and documents. While insightful, such work does not enable us to gain deep insights into a single model. Closer to our work, Rosset et al.~\cite{DBLP:conf/sigir/RossetMXCST19} employ axioms to generate artifical documents for the training of neural models and the regularization of the loss function. In contrast, we employ axioms to \emph{analyze} retrieval models.

Another direction of research has been the development of interpretation tools such as DeepSHAP~\cite{FernandoSA19SIGIR} and LIRME~\cite{VermaG19SIGIR} that aim to generate \emph{local} explanations for neural IR models. Recently, in particular \bert{} (due to its successes across a wide range of tasks and domains) has become the focus of analysis---not within IR though. While approaches like~\cite{ClarkKLD2019ArXiv} explore the attention values generated by the model's attention layers, Tenney et al.~\cite{TenneyDP2019ACL} argue that \bert{} is re-discovering classical NLP pipeline approaches in its layers, \textit{``in an interpretable and localizable way"}, essentially repeating traditional NLP steps in a similar order as an expert would do, with steps like POS tagging, parsing, NER and coreference resolution happening within its layers in the expected order. Finally, Niven et al.~\cite{NivenK19ACL} raise some critical points about~\bert{}, arguing that it only \textit{``exploits spurious statistical cues in the dataset"}; they showcase this by creating adversarial datasets that can significantly harm its performance.

\section{Diagnostic Datasets}\label{sec:diag_datasets}

The usage of diagnostic datasets as a means to analyse neural models is common in NLP, e.g.~\cite{weston2015towards,jia2017adversarial,wang2018glue} as there are a large number of fine-grained linguistic tasks (anaphora resolution, entailment, negation, etc.) that datasets can be created for with relative ease. In contrast, in IR the central notion is relevance and although we know that it can be decomposed into various types (topical, situational, etc.) of relevance~\cite{saracevic1996relevance}, we have no easy way of creating datasets for each of these---it remains a time-intensive and expensive task. This also explains why corpora such as \robust{} remain useful and in use for such a long time. Instead, like Rennings et al.~\cite{RenningsMH19ECIR} we turn to the axiomatic approach to IR and create diagnostic datasets---one for each of our chosen retrieval heuristics. It has been shown that, generally, retrieval functions that fulfil these heuristics achieve a greater effectiveness than those that do not. In contrast to~\cite{RenningsMH19ECIR} which restricted itself to four term matching axioms, we explore a wider range of axioms, covering term frequency, document length, lower-bounding term frequency, semantic term matching and term proximity constraints. In total, we explore 9 axioms, all of which are listed in Table~\ref{tab:heuristics} with a short informal description of their main assumption of what a sensible retrieval function should fulfil. We note that this covers most of the term-matching and semantic-matching axioms that have been proposed. We have eliminated a small number from our work as we do not consider them relevant to \bert{} (e.g., those designed for pseudo-relevance feedback~\cite{clinchant2011document}).

As the axiomatic approach to IR has been designed to \emph{analytically} analyse retrieval functions, in their original version they assume very specific artificial query and document setups. As a concrete example, let us consider axiom \STMCI{}~\cite{Fang06SIGIR}. It is defined as follows: \emph{given a single-term query $Q=\{q\}$ and two single-term documents $D_1=\{d_1\}$, $D_2=\{d_2\}$ where $d_1\ne d_2 \ne q$, the retrieval score of $D_1$ should be higher than $D_2$ if the semantic similarity between $q$ and $d_1$ is higher than that between $q$ and $d_2$}. This description is sufficient to mathematically analyse classic retrieval functions, but not suitable for models with more than a handful of parameters. We thus turn to the creation of datasets that \emph{exclusively} contain instances of query/documents (for \STMCI{} an instance is a triple, consisting of one query and two documents) that satisfy a particular axiom. As single-term queries and documents offer no realistic test bed, we \emph{extend} (moving beyond single-term queries and documents) and \emph{relax} (moving beyond strict requirements such as equal document length) the axioms in order to extract instances from existing datasets that fulfil the requirements of the extended and relaxed axiom. Importantly, this process requires no relevance judgements---we can simply scan all possible triples in the corpus (consisting of queries and documents) and add those to our diagnostic dataset that fulfil our requirements. We then score each query/document pair with our \bert{} model\footnote{Note, that scoring each document independently for each query is an architectural choice, there are neural architectures that take a query/doc/doc triplet as input and output a preference score.} and determine whether the score order of the documents is in line with the axiom. If it is, we consider our model to have classified this instance correctly. 

While Table~\ref{tab:heuristics} provides an informal overview of each heuristic, we now formally describe each one in more detail. Due to the space limitations, we focus on a mathematical notation which is rather brief. For completeness, we first state the original axiom and then outline how we extend and relax it in order to create a diagnostic dataset from it. For axioms \TFCI{}, \TFCII{}, \LNCII{} and \MTDC{} we follow the process described in~\cite{RenningsMH19ECIR}. We make use of the following notation: $Q$ is a query and consists of terms $q_1, q_2, ...$; $D_i$ is a document of length $|D_i|$ containing terms $d_{i_1}, d_{i_2}, ...$; the count of term $w$ in document $D$ is $c(w,D)$; lastly, $S(Q,D)$ is the retrieval score the model assigns to $D$ for a given $Q$. Apart from the proximity heuristic \TP{}, the remaining heuristics are based on the bag-of-word assumption, i.e., the order of terms in the query and documents do not matter.

\paragraph{\TFCI{}---Original} Assume $Q=\{q\}$ and $|D_1|=|D_2|$. If $c(q,D_1)>c(q,D_2)$, then $S(Q,D_1)>S(Q,D_2)$. \smallskip
\paragraph{\TFCI{}---Adapted} In order to extract query/document/document triples from actual corpora, we need to consider multi-term queries and document pairs of approximately the same length. Let $Q=\{q_1,q_2,..,q_{|Q|}\}$ and $|D_1|-|D_2|\leq \textit{abs}(\delta)$. $S(Q,D_1)>S(Q,D_2)$ holds, when $D_1$ has at least the same query term count as $D_2$ for all but one query term (and for this term $D_1$'s count is higher), i.e., $c(q_i,D_1)\geq c(q_i,D_2)\ \forall q_i \in Q$ and $\sum_{q_i \in Q} c(q_i,D_1) > \sum_{q_i \in Q}c(q_i,D_2)$.

\paragraph{\TFCII{}---Original} Assume $Q=\{q\}$ and $|D_1|=|D_2|=|D_3|$. If $c(q, D_1)>0$, $c(q, D_2)-c(q, D_1)=1$ and $c(q, D_3)-c(q, D_2)=1$, then $S(Q, D_2)-S(Q,D_1) > S(Q,D_3)-S(Q,D_2)$.\smallskip

\paragraph{\TFCII{}---Adapted} Analogous to \TFCI{}, queries can contain multiple terms and documents only have to have approximately the same length. Let $Q=\{q_1,q_2,..,q_{|Q|}\}$ and $max_{D_i,D_j\in \{D_1, D_2, D_3\}}(|D_i|-|D_j|\leq \textit{abs}(\delta))$. If every document contains at least one query term, and $D_3$ has more query terms than $D_2$ and $D_2$ has more query terms than $D_1$, and the difference of query terms count between $D_2$ and $D_1$ should be the same as between $D_3$ and $D_2$, for all query terms, i.e. $\sum_{q\in Q}c(q, D_3)>\sum_{q\in Q}c(q, D_2)>\sum_{q\in Q}c(q, D_1)>0$ and $c(q, D_2)-c(q,D_1)=c(q, D_3)-c(q,D_2) \forall q \in Q$, then $S(Q, D_2)-S(Q,D_1) > S(Q,D_3)-S(Q,D_2)$.

\paragraph{\MTDC{}---Original} Let $Q=\{q_1, q_2\}$, $|D_1|=|D_2|$, $c(q_1, D_1)=c(q_2, D_2)$ and $c(q_2, D_1)=c(q_1, D_2)$. If $idf(q_1)\geq idf(q_2)$ and $c(q_1, D_1) > c(q_1, D_2)$, then $S(Q, D_1)\geq S(Q, D_2)$.\smallskip 

\paragraph{\MTDC{}---Adapted} Again, Let $Q$ contain multiple terms and $|D_1|-|D_2|\leq \textit{abs}(\delta)$. $D_1, D_2$ also differ in at least one query term count ($\exists q_i \in Q, \text{ such that } c(q_i, D_1)\neq c(q_i, D_2)$). If, for all query term pairs the conditions hold that $c(q_i, D_1)\neq c(q_j, D_j)$, $idf(q_i) \geq idf(q_j)$, $c(q_i, D_1) = c(q_j, D_2)$, $c(q_j, D_1)=c(q_i, D_2)$, $c(q_i, D_1)>c(q_i, D_2)$ and $c(q_i, Q) \geq c(q_j, Q)$, then $S(Q, D_1) \geq S(Q, D_2)$.

\paragraph{\LNCI{}---Original}  Let $Q$ be a query and $D_1, D_2$ be two documents. If for some $q'\notin Q$, $c(q', D_2)=c(q', D_1)+1$ and for any $q\in Q$ $c(q,D_2)=c(q, D_1)$, then $S(Q, D_1)\geq S(Q, D_2)$. \smallskip
\paragraph{\LNCI{}---Adapted} The axiom can be used with no adaptation.

\paragraph{\LNCII{}---Original} Let $Q$ be a query. $\forall k >1$, if $D_1$ and $D_2$ are two documents such that $|D_1|=k\cdot{}|D_2|$, and $\forall q \in Q, c(q, D_1) = k\cdot{}c(q, D_2)$, then $S(Q, D_1)\geq S(Q, D_2)$. \smallskip
\paragraph{\LNCII{}---Adapted} The axiom can be used with no adaptation.

\paragraph{\TP{}---Original} Let $Q=\{q_1, q_2, ... q_{|Q|}\}$ be a query and $D'$ a document generated by switching the position of query terms in $D$. Let $\sigma(Q, D)$ be a function that measures the distance of query terms $q_i \in Q$ inside a document $D$. If $\sigma(Q, D) > \sigma(Q, D')$, then $S(Q, D) < S(Q, D')$. \smallskip
\paragraph{\TP{}---Adapted} Let $\Gamma(D, Q) = min_{(q_1,q_2\in Q\cap D,q_1\neq q_2)} {Dis(q_1, q_2; D)}$ be a function that computes the minimum distance between every pair of query terms in $D$. If $\Gamma(D_1, Q) < \Gamma(D_2, Q)$, then $S(Q, D_1) > S(Q, D_2)$.

\vspace{0.25cm}
For the following semantic axioms, let us define the function $\sigma(t_1, t_2)$ as the cosine distance between the embeddings of terms $t_1$ and $t_2$. We also define $\sigma'(T_1, T_2)$, where $T$ can be either a document $D$ or a query $Q$, as an extension to $\sigma$, defined by $\sigma'(T_1, T_2) = cos(\frac{\sum_{i\in T_1}t_i}{|T_1|}, \frac{\sum_{i\in T_2}t_i}{|T_2|})$, the cosine distance between the average term embeddings for each document.

\paragraph{\STMCI{}---Original} Let $Q=\{q\}$ be a one-term query, $D_1=\{d_{1_1}\}$ and $D_2=\{d_{2_1}\}$ be two single term documents, such that $d_{1_1}\neq d_{2_1}$, $q\neq d{1_1}$ and $q\neq d_{2_1}$. If $\sigma(q, d_{1_1}) > \sigma(q, d_{2_1})$, then $S(Q, D_1) > S(Q, D_2)$.\smallskip

\paragraph{\STMCI{}---Adapted} We allow $D_1$ and $D_2$ to be arbitrarily long, covering the same number of query terms (i.e. $|D_1\bigcap Q| = |D_2\bigcap Q|$). Assume $\{D_i\}-\{Q\}$ be the document $D_i$ without query terms, If $\sigma'(\{D_1\}-\{Q\}, Q) > \sigma'(\{D_2\}-\{Q\}, Q)$, then $S(Q, D_1) > S(Q, D_2)$.

\paragraph{\STMCII{}---Original} Let $Q=\{q\}$ be a one-term query and $d$ a non-query term such that $\sigma(d, q)>0$. If $D_1=\{q\}$ and $|D_2|=k, (k\geq1)$, composed entirely of $d$'s (i.e.v$c(d, D_2)=k$), then $S(Q, D_1) \ge S(Q, D_2)$. \smallskip
\paragraph{\STMCII{}---Adapted} We allow $Q$ to be a multiple query term, $D_1$ to contain non-query terms and $D_2$ to contain query terms. If $\sum_{t_i, t_i \notin Q} c(t_i, D_2)>\sum_{q_i \in Q}c(q_1, D_1)>0$, $\sigma'(\{D_1\}-\{Q\}, \{D_2\}-\{Q\}) > \delta$ then $S(Q, D_1) \ge S(Q, D_2)$.
 
\paragraph{\STMCIII{}---Original} Let $Q=\{q_1, q_2\}$ be a two-term query and $d$ a non-query term such that $\sigma(d, q_2)>0$. If $|D_1|=|D_2|>1$, $c(q_1, D_1)=|D_1|$, $c(q1, D_2) = |D_2|-1$ and $c(d, D_2)=1$, then $S(Q, D_1) \leq S(Q, D_2)$.\smallskip

\paragraph{\STMCIII{}---Adapted} Let $D_1$ and $D_2$ be two arbitrary long documents that covers the same number of query terms (i.e. $|D_1\bigcap Q| = |D_2\bigcap Q|$). If $|D_1|-|D_2| \leq \textit{abs}(\delta)$, $\sum_{q_i \in Q} c(q_i,D_1) > \sum_{q_i \in Q}c(q_i,D_2)$ and $\sigma'(\{D_2\}-\{Q\}, Q)>\sigma'(\{D_1\}-\{Q\}, Q)$, then $S(Q, D_1)>S(Q, D_2)$.

\begin{table}[!htb]
    \centering
    \small
    \caption{Overview of retrieval heuristics employed in our work. The diagnostic datasets for heuristics marked with a \colorbox{aliceblue}{blue background} were first discussed in~\cite{RenningsMH19ECIR}. The naming of the heuristics is largely taken from the papers proposing them.}
    \label{tab:heuristics}
    \begin{tabular}{lp{10cm}}
    \toprule
     \textbf{Heuristic} & \textbf{Informal description}\\
     \midrule
     \multicolumn{2}{l}{\textit{\textcolor{gray}{Term frequency constraints}}} \\
     \rowcolor{aliceblue}
     \TFCI~\cite{FangTZ04SIGIR}     & The more occurrences of a query term a document has, the higher its retrieval score. \\
     \rowcolor{aliceblue}
     \TFCII~\cite{FangTZ04SIGIR}    & The increase in retrieval score of a document gets smaller as the absolute query term frequency increases. \\
     \rowcolor{aliceblue}
     \MTDC~\cite{FangTZ04SIGIR,shi2005gravitation} & The more discriminating query terms (i.e., those with high IDF value) a document contains, the higher its retrieval score. \\
     
     \midrule
    \multicolumn{2}{l}{\textit{\textcolor{gray}{Length normalization constraints}}} \\
     \LNCI~\cite{FangTZ04SIGIR}     & The retrieval score of a document decreases as terms not appearing in the query are added. \\
     \rowcolor{aliceblue}
     \LNCII~\cite{FangTZ04SIGIR}    & A document that is duplicated does not have a lower retrieval score than the original document. \\
     

     \midrule
    \multicolumn{2}{l}{\textit{\textcolor{gray}{Semantic term matching constraints}}} \\
    \STMCI{}~\cite{Fang06SIGIR}     & A document's retrieval score increases as it contains terms that are more semantically related to the query terms. \\
    \STMCII{}~\cite{Fang06SIGIR}    & The document terms that are a syntactic match to the query terms contribute at least as much to the document's retrieval score as the semantically related terms.\\
    \STMCIII{}~\cite{Fang06SIGIR}   & A document's retrieval score increases as it contains more terms that are semantically related to \emph{different} query terms.  \\
    
    \midrule
    \multicolumn{2}{l}{\textit{\textcolor{gray}{Term proximity constraint}}} \\
    \TP{}~\cite{tao2007exploration} & A document's retrieval score increases as the query terms appearing in it appear in closer proximity. \\
     \bottomrule
    \end{tabular}
\end{table}

\section{Experiments}\label{sec:experiments}


We create diagnostic datasets for each of these axioms by extracting instances of queries and documents that already exist in the dataset. In this section, we explain how these datasets were generated and how we employed them to evaluate \bert{}.

\subsection{TREC 2019 Deep Learning Track}\label{sec:dataset}
In order to extract diagnostic datasets, we used the corpus and queries for the Document Ranking Task from the TREC 2019 Deep Learning track\footnote{\url{https://microsoft.github.io/TREC-2019-Deep-Learning/}}. This is the only publicly available ad-hoc retrieval dataset that was built specifically for the training of deep neural models, with 3,213,835 web documents and 372,206 queries (367,013 queries in the training set and 5,193 in the development set). The queries and documents, while stripped of HTML elements, are not necessarily well-formed as seen in the following examples from the training set:

\begin{itemize}
    \item \texttt{what is a flail chest}\smallskip
    \item \texttt{a constitution is best described as a(n) \_\_\_\_\_\_\_\_\_\_.}
    \item \texttt{	)what was the immediate impact of the success of the manhattan project?}
\end{itemize}

The queries consist on average of $5.89(\pm 2.51)$ words while documents consist on average of $1084.88(\pm 2324.22)$ words. 

Most often, one relevant document exists per query ($1.04$ relevant documents on average). These relevance judgements were made by human judges on a passage-level: if a passage within a document is relevant, the document is considered relevant. Unlike other TREC datasets, like \robust{}, there is no topic description or topic narrative.

At the time of this writing, the relevance judgements for the test queries were not available. Therefore, we split the development queries further, in a new \textit{dev} and \textit{test} dataset, in a $70\%-30\%$ fashion. In the rest of this paper, when we refer to the \textit{test} or \textit{dev} dataset, we are referring to this split. The \textit{train} split remains the same as the original dataset.

\subsection{Retrieval and Ranking}
We begin by indexing the document collection using the Indri toolkit\footnote{\url{https://www.lemurproject.org/indri.php}}, and retrieve the top-100 results using a traditional retrieval model with just one hyperparameter, namely, QL, with Indri's default setting (Dirichlet smoothing~\cite{zhai2017study} and $\mu=2500$). Finally, for \bert{}, we employ Hugging Face's library\footnote{\url{https://github.com/huggingface/transformers}} of Distil\bert{}~\cite{sanh2019distilbert}, a distilled version of the original \bert{} model, with fewer parameters (66 million instead of 340 million), and thus more efficient to train, but with very similar results.

We fine-tuned our \bert{}\footnote{Code for fine-tuning Distil\bert{} and generating the diagnostic datasets is available at \url{https://github.com/ArthurCamara/bert-axioms}} model with 10 negative samples for each positive sample from the training dataset, randomly picked from the top-100 retrieved from QL. We set the maximum input length to 512 tokens. 
For fine-tuning we used the sequence classification model. It is implemented by adding a fully-connected layer on top of the \texttt{[CLS]} token embedding, which is the specific output token of the \bert{} model that our fine-tuning is based upon.

Given the limitation of \bert{} regarding the maximum number of tokens, we limited the document length to its first 512 tokens, though we note that alternative approaches exist (e.g., in~\cite{YangZL2019Arxiv} the \bert{} scores across a document's passages/sentences are aggregated). In Table~\ref{tab:axioms1}, we report the retrieval effectiveness in terms of nDCG and MRR for the documents limited to 512 tokens. We rerank the top-100 retrieved documents based on its first 512 tokens. It is clear that \bert{} is vastly superior to QL with a 40\% improvement in $nDCG$ and 25\% improvement in $MRR$. 



\begin{table}
\caption{Overview of the number of instances in each diagnostic dataset (row~I), the number of instances within each diagnostic dataset that contain a relevant document (row~II) and the fraction of instances among all of row~II where the order of the documents according to the axiom is in line with the relevance judgments. \LNCII{} is based on new documents, thus, it does not have a fraction of agreement}
\label{tab:dataset}
\resizebox{\textwidth}{!}{%
\begin{tabular}{lrrrrrrrrrrr}
\toprule
 & \TFCI{} & \TFCII{} & \MTDC{} & \LNCI{} & \LNCII{}   & \TP{} & \STMCI{} & \STMCII{} & \STMCIII{}  \\
 \midrule
Diagnostic dataset size & 119,690 & 10,682 & 13,871 & 14,481,949 & 7452 &  3,010,246 & 319,579 & 7,321,319 &  217,104\\
Instances with a \emph{relevant} document & 1,416 & 17 & 11 & 138,399 & 82  &  20,559 & 19,666 & 70,829 & 1,626 \\ 
Fraction of instances agreeing with relevance & 0.91 & 0.29 & 0.82 & 0.50 & - &  0.18 & 0.44 & 0.63 & 0.35  \\ 
\bottomrule
\end{tabular}
}
\end{table}

\subsection{Diagnostic Datasets}\label{sec:datasets}
Given the adapted axioms defined in Section~\ref{sec:diag_datasets}, we now proceed on describing how to extract actual datasets from our corpus.

\paragraph{\TFCI{}, \TFCII{}, \MTDC{}, \LNCI{}}We add tuples of queries and documents $\{q, d_i, d_j\}$ (or $\{q, d_i, d_j, d_k\}$ for \TFCII{}) for every possible pair of documents $\{d_i, d_j\}$ in the top 100 retrieved by QL that follow the assumptions from Section~\ref{sec:diag_datasets}, with $\delta=10$. We also compute \texttt{IDF} for \MTDC{} on the complete corpus of documents, tokenized by WordPiece~\cite{DBLP:journals/corr/WuSCLNMKCGMKSJL16}.
\vspace{-10pt}
\paragraph{\LNCII{}}We create a new dataset, appending the document to itself $k\in \mathbb{Z}$ times until we reach up to 512 tokens\footnote{Note that we only append the document to itself if the final size does not exceed 512.}. In contrast to Rennings et al.~\cite{RenningsMH19ECIR} we only perform this document duplication for our test set, i.e., \bert{} does not ``see'' this type of duplication during its training phase. On average, the documents were multiplied $k=2.6408\pm 1.603$ times, with a median of $k=2$.
\vspace{-10pt}
\paragraph{\TP{}} We simply add to our dataset every pair of documents $\{q, d_i, d_j\}$ in the top 100 retrieved documents by QL for a given topic that follow the stated \TP{} assumptions.
\vspace{-10pt}
\paragraph{\STMCI{},\STMCII{}, \STMCIII{}} We define $\sigma$ as the cosine distance between the embeddings of the terms and $\sigma'$ as the cosine distance between the average embeddings. We trained the embeddings using GLoVe~\cite{PenningtonSM14@EMNLP} on the entire corpus. For \STMCIII{}, we set $\delta=0.2$.

\subsection{Results}
In Table~\ref{tab:dataset} we list the number of diagnostic instances we created for each diagnostic dataset. In addition, we also performed a sanity check on the extent to which the document order determined by each axiom corresponds to the relevance judgements. Although only a small set of instances from each diagnostic dataset contains a document with a relevant document (row II in Table~\ref{tab:dataset}) we already see a trend: apart from \TFCI{} and \MTDC{} where 91\% and 82\% of the diagnostic instances have an agreement between axiomatic ordering and relevance ordering, the remaining axioms are actually not in line with the relevance ordering for most of the instances. This is a first indication that we have to consider an alternative set of axioms, better fit for such a corpus, in future work.

In Table~\ref{tab:axioms1} we report the fraction of instances both QL and \bert{} fulfil for each diagnostic dataset. As expected, QL \emph{correctly} (as per the axiom) ranks the document pairs or triples most of the time, with the only outlier being \TP{}, where QL performs essentially random---again, not a surprise given that QL is a bag-of-words model. In contrast---with the exception of \LNCII{}, where \bert{}'s ranking is essentially the opposite of what the \LNCII{} axiom considers correct (with only 6\% of the instances ranked correctly)---\bert{} has not learnt anything that is related to the axioms as the fraction of correctly ranked instances hovers around 50\% (which is essentially randomly picking a document order). Despite this lack of axiomatic fulfilment, \bert{} clearly outperforms QL, indicating that the existing axioms are not suitable to analyse \bert{}.

The reverse ranking \bert{} proposes for nearly all of the \LNCII{} instances can be explained by the way the axiom is phrased. It is designed to avoid over penalising documents, and thus a duplicated document should always have a retrieval score that is not lower than the original document. The opposite argument though could be made too (and \bert{} ranks accordingly), that a duplicated document should not yield a higher score than the original document as it does not contain novel/additional information. As we did not provide \LNCII{} instances in the training set, \bert{} is not able to rank according to the axiom, in line with the findings of other neural approaches as shown by Rennings et al.~\cite{RenningsMH19ECIR}.

Finally, we observe that, counter-intuitively, \bert{} does not show a performance better than QL for semantic term matching constraints. For instance, one may expect that \bert{} would fare quite well on \STMCI{}, given its semantic nature. However, our results indicate that \bert{} is actually considers term matching as one of its key features. In order to further explore this tension between semantic and syntactic term matching, we split the queries in our test set by their term overlap between the query and the relevant document (if several relevant documents exist for a query, we randomly picked one of them). If a query/document has no (or little) term overlap, we consider this as a semantic match.

 

\begin{table}
\caption{Overview of the retrieval effectiveness (nDCG columns) and the fraction of diagnostic dataset instances each model ranks correctly.}
\label{tab:axioms1}
\resizebox{\textwidth}{!}{%
\begin{tabular}{lrr|rrrrrrrrrr}
\toprule
 &  nDCG & MRR & \TFCI{} & \TFCII{} & \MTDC{} & \LNCI{} & \LNCII{} & \TP{}  & \STMCI{} & \STMCII{} & \STMCIII{} \\ 
\midrule
QL &  0.2627 & 0.3633 & \textbf{0.99} & \textbf{0.70} & \textbf{0.88} &\textbf{ 0.50} & \textbf{1.00} & 0.39  & 0.49 & \textbf{0.70} &  \textbf{0.70} \\
 
DistilBERT &  \textbf{0.3633} &  \textbf{0.4537} & 0.61 & 0.39 & 0.51 & \textbf{0.50} & 0.00 & \textbf{0.41} & \textbf{0.50} & 0.51 & 0.51 \\

\bottomrule
\end{tabular}
}

\end{table}%

\begin{figure}
    \centering
    \begin{minipage}{0.5\textwidth}
        \centering
        \includegraphics[width=\textwidth]{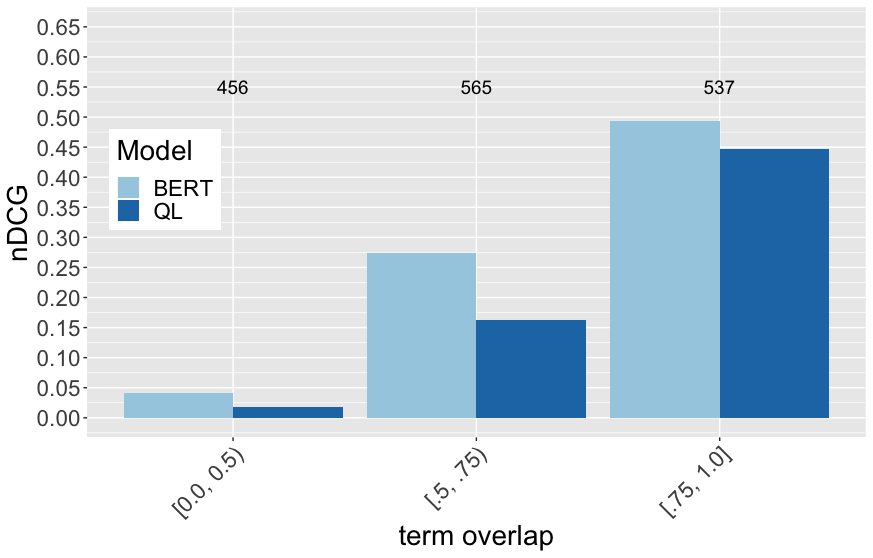}
        \label{fig:term_overlap1}
    \end{minipage}\hfill
    \begin{minipage}{0.5\textwidth}
        \centering
        \includegraphics[width=\textwidth]{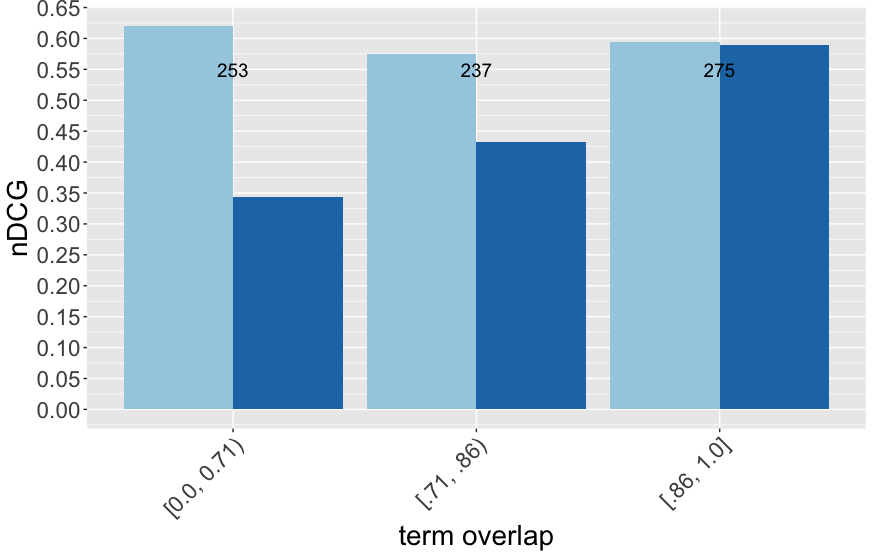} 
        \label{fig:term_overlap2}
    \end{minipage}
    \caption{The test queries are split into three sets, depending on the fraction of term overlap between the query and its corresponding relevant document. On the left, we plot all queries, on the right only those queries for which the relevant document appears in the top-100 ranked documents of the QL ranking.}
    \label{fig:term_overlap}
\end{figure}

 The results of this query split can be found in Figure~\ref{fig:term_overlap}. We split the query set roughly into three equally sized parts based on the fraction of query terms appearing in the relevant document (as an example, if a query/document pair has a fraction of 0.5, half of all query terms appear in the document). We report results for all queries (Figure~\ref{fig:term_overlap} (left)), as well as only those where the relevant document appears in the top-100 QL ranking (Figure~\ref{fig:term_overlap} (right)). We find that \bert{} outperforms QL across all three splits, indicating that \bert{} is indeed able to pick up the importance of syntactic term matching. At the same time, as expected, \bert{} is performing significantly better than QL for queries that require a large amount of semantic matching. That brings into question on why, then, the axiomatic performance across our semantic axiomatic datasets does not reflect that. One hypothesis is that the semantic similarity we measure (based on context-free word embeddings) is different to the semantic similarity measured via contextual word embeddings. This in itself is an interesting avenue for future work, since it brings a new question on \textit{what} that semantic relationship is, and how to accurately measure it.

\vspace{-1cm}
\section{Discussion \& Conclusion}\label{sec:conclusion}

In this paper, we set out to analyze \bert{} with the help of the recently proposed \emph{diagnostic datasets for IR based on retrieval heuristics} approach~\cite{RenningsMH19ECIR}. We expected \bert{} to perform better at fulfilling some of the proposed semantic axioms. Instead, we have shown that \bert{}, while significantly better than traditional models for ad-hoc retrieval, does not fulfil most retrieval heuristics, created by IR experts, that are supposed to produce better results for ad-hoc retrieval models. We argue that based on these results, the axioms are not suitable to analyse \bert{} and it is an open question what type of axioms would be able to capture some performance aspects of \bert{} and related models. In fact, how to arrive at those additional axioms, based on the knowledge we have now gained about \bert{} is in itself an open question. 

\vspace{\baselineskip}

\noindent\textbf{Acknowledgement}. This research has been supported by NWO project SearchX (639.022.722).


\bibliographystyle{splncs04}
\bibliography{references}

\end{document}